# Super-resonances in microspheres: extreme effects in field localization

Zengbo Wang[1,*], Boris Luk'yanchuk[2,3,*], Liyang Yue[1], Ramón Paniagua-Domínguez[4], Bing Yan[1], James Monks[1], Oleg V. Minin[5], Igor V. Minin[6], Sumei Huang[7] and Andrey A. Fedyanin[3]

[1] School of Electronic Engineering, Bangor University, Dean Street, Bangor, Gwynedd, LL57 1UT, UK
[2] Division of Physics and Applied Physics, School of Physical and Mathematical Sciences, Nanyang Technological University, Singapore 637371, Singapore
[3] Faculty of Physics, Lomonosov Moscow State University, Moscow 119991, Russia
[4] Institute of Materials Research and Engineering, Agency for Science, Technology and Research, Singapore, Singapore.
[5] National Research Tomsk State University, Lenin Ave., 36, Tomsk, 634050, Russia
[6] National Research Tomsk Polytechnic University, Lenin Ave., 30, Tomsk, 634050, Russia
[7] Engineering Research center for Nanophotonics & Advanced Instrument, Ministry of Education, School of Physics and material Science, East China Normal Univesity, North Zhongshuan Rd. 3663, Shanghai 200062, PR China

Correpondence: z.wang@bangor.ac.uk and lukiyanchuk@nanolab.phys.msu.ru

**Abstract**

We reveal the existence of optical, super-resonance modes supported by dielectric microspheres. These modes, with field-intensity enhancement factors on the order of $10^4$–$10^5$, can be directly obtained from analytical calculations. In contrast to usual optical resonances, which are related to the poles of the electric and magnetic scattering amplitudes, **super-resonance modes are related to the poles of the internal field coefficients**, obtained for specific values of the size parameter. We also reveal the connection of these super-resonances in the generation of magnetic nanojets and of giant magnetic fields in particles with high refractive index.

## 1. Introduction

This year sees the 111$^{th}$ anniversary of Gustav Mie's publication[1], in which he presented a complete solution of Maxwell's equations for light scattering of a linearly polarized electromagnetic plane wave by a homogeneous, isotropic sphere. Although Mie was not the first to obtain such solution (see a historical picture in Ref. [2]), he can be considered as one of the fathers of resonant light scattering, a phenomenon that nowadays plays a pivotal role in modern nanophotonics[3]. Indeed, the rapid growth in the publication activity related to Mie theory, as seen in Fig. 1, can be connected with the successive understanding of numerous novel physical phenomena observed in the past three decades, phenomena that were encrypted within the theory and could be explained thanks to it.

As examples of these phenomena, one could mention the near field focusing effect[4-9] produced by a sphere with a big size parameter (which describes the ratio between the physical size of the scatter and the illuminating wavelength). This effect was subsequently referred to as the "*photonic nanojet*" [10-14]. One could also mention the plasmonic resonances of small metallic particles and the Fano resonances in both plasmonic materials and metamaterials[15-17]. The formation of nanovortices[18-21], in which the local wave vectors can greatly exceed the wave vector of incident radiation, the nanoscopy imaging[22,23,14] or the existence of "*anapole*" states – non-radiating charge-current configurations excited in dielectric nanoparticles[24-27] – could be also mentioned as examples of new phenomena found with the help of Mie theory. In this regard, a separate mention should be made of the whole range of novel effects found and explained, partially thanks to Mie theory, in nanoparticles made of materials with a high refractive index. Of particular importance are the excitation of optical resonances with magnetic multipolar moments[28, 29], which provide "magnetic light"[30, 31], and the possibility to obtain strongly directional light scattering via angular interference[32-35]. It should be emphasized, once again, that all these effects were somehow encrypted within the formulas written 150 years ago, and they were just waiting for somebody to decrypt them.

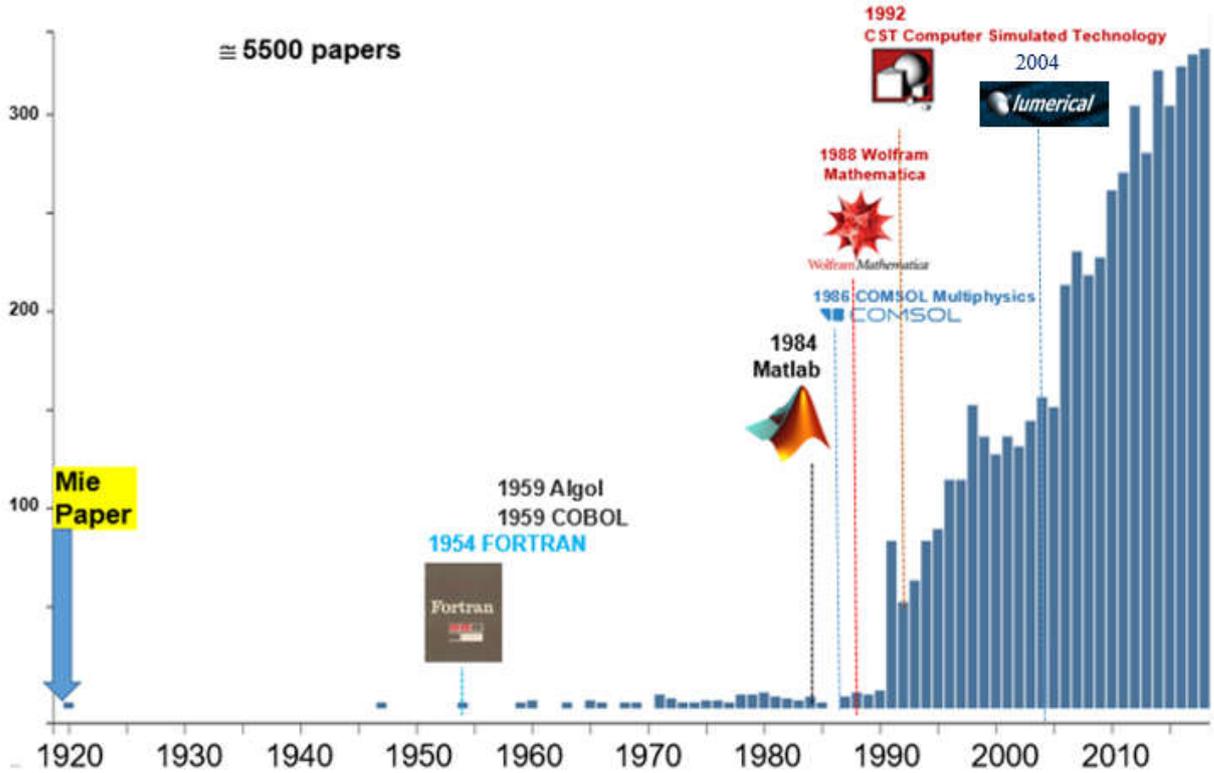

Fig. 1. Number of publications on the Mie theory according to Web of Science. Dates of important software are inserted. The first peak in publications in 1991 could be linked to the advent of plasmonics.

The power of Mie theory is that it accurately describes scattering phenomena all the way from the Rayleigh regime to that of geometrical optics. In particular, far field scattering effects[36-39], including optical resonances[39] and scattering diagrams, have been extensively analysed in many different works for particles with small size parameter $q \cong 1$ (typically with a limited number of multipoles, e.g. $\ell_{max} \leq 4$). This corresponds to nanometre-scale particles for light in the visible range. On the other side of the range of size parameters, typically $q > 100$, the realm of geometrical optics, working with millimetre-to-centimetre scale objects, has also been extensively studied. In this case, the majority of calculations relies on the geometrical optics approximation, which may provide very good agreement with the full solution[40]. Between the scales of "*nano-optics*" ($q \cong 1$) and "*geometrical optics*" ($q > 100$) there is the intermediate range of size parameters ($q \cong 10$), for which the above mentioned "*photonic nanojet*" effect was found. For visible light, this range of size parameters correspond to micrometer-scaled objects (typically from a few to few tens micrometres).

In this work, we describe a novel effect found in this intermediate ($q \cong 10$) range of size parameters, namely, the excitation of super-resonance modes in dielectric spheres. These modes are characterized by very large enhancements ($10^4$–$10^5$) of the near field-intensities, and in particular of the magnetic one. We relate these resonances with the poles in the internal field coefficients in the Mie theory.

## 2. Results and discussion

Mie theory expresses the scattered fields as a superposition of partial waves in terms of spherical harmonics[36-38]. According the theory, the total scattering efficiency is presented as a sum of partial scattering efficiencies:

$$Q_{sca} = \sum_{\ell=1}^{\infty} \left( Q_\ell^{(e)} + Q_\ell^{(m)} \right), \qquad Q_\ell^{(e)} = \frac{2(2\ell+1)}{q^2} |a_\ell|^2, \qquad Q_\ell^{(m)} = \frac{2(2\ell+1)}{q^2} |b_\ell|^2, \qquad (1)$$

where each partial efficiency $Q_\ell^{(e)}$ and $Q_\ell^{(m)}$ describe the scattering associated, respectively, to the $\ell$-th order electric and magnetic multipolar polarizability (dipolar, $\ell = 1$, quadrupolear, $\ell = 2$, octupolear, $\ell = 3$, etc.). Thus, Mie theory allows a clear understanding of the origin of different resonances observed in the far-field scattering from a particle. In the following, we will discuss only transparent dielectrics, with $\text{Im}\,\varepsilon = 0$, so $Q_{ext} = Q_{sca}$. The electric, $a_\ell$, and magnetic, $b_\ell$, scattering amplitudes for nonmagnetic materials ($\mu = 1$), and dielectric permittivity $\varepsilon = n^2$ ($n$ being the refractive index of the particle material) are given by:

$$a_\ell = \frac{\mathfrak{R}_\ell^{(a)}}{\mathfrak{R}_\ell^{(a)} + i\,\mathfrak{I}_\ell^{(a)}}, \qquad b_\ell = \frac{\mathfrak{R}_\ell^{(b)}}{\mathfrak{R}_\ell^{(b)} + i\,\mathfrak{I}_\ell^{(b)}}, \qquad (2)$$

where $\mathfrak{R}_\ell$ and $\mathfrak{I}_\ell$ functions are defined as follows:

$$\mathfrak{R}_\ell^{(a)} = n\psi_\ell'(q)\psi_\ell(nq) - \psi_\ell(q)\psi_\ell'(nq), \quad \mathfrak{I}_\ell^{(a)} = n\chi_\ell'(q)\psi_\ell(nq) - \chi_\ell(q)\psi_\ell'(nq), \qquad (3)$$

$$\mathfrak{R}_\ell^{(b)} = n\psi_\ell'(nq)\psi_\ell(q) - \psi_\ell(nq)\psi_\ell'(q), \quad \mathfrak{I}_\ell^{(b)} = n\chi_\ell(q)\psi_\ell'(nq) - \psi_\ell(nq)\chi_\ell'(q). \qquad (4)$$

Here, $\psi_\ell(z) = \sqrt{\frac{\pi z}{2}}\,J_{\ell+\frac{1}{2}}(z)$, $\chi_\ell(z) = \sqrt{\frac{\pi z}{2}}\,N_{\ell+\frac{1}{2}}(z)$, where $J_{\ell+\frac{1}{2}}(z)$ and $N_{\ell+\frac{1}{2}}(z)$ are the Bessel and Neumann functions. The radius of the particle $R$ enters in this theory through the dimensionless size parameter $q = \omega R/c = 2\pi R/\lambda$, where $\omega$ is the angular frequency, $c$ the speed of light, and $\lambda$ the radiation wavelength in vacuum. The prime in formulas (3), (4) indicates differentiation with respect to the argument of the function, e.g. $\psi_\ell'(z) \equiv d\psi_\ell(z)/dz$.

In an analogous way, Mie theory allows one to write the electric and magnetic fields inside the particle through the internal field amplitudes, given by:

$$c_\ell = \frac{i\,n}{\mathfrak{R}_\ell^{(a)} + i\,\mathfrak{I}_\ell^{(a)}}, \quad d_\ell = \frac{i\,n}{\mathfrak{R}_\ell^{(b)} + i\,\mathfrak{I}_\ell^{(b)}}. \qquad (5)$$

Although the denominators of these amplitudes are the same as in the scattering coefficients, $a_\ell$ and $b_\ell$ in (2), which means that the spectral position of these resonances are close, the numerators of (5) never vanish. As a result, the values of amplitudes $|c_\ell|^2$ and $|d_\ell|^2$ are not bounded, unlike $|a_\ell|^2$ and $|b_\ell|^2$ in (2) which can never exceed unity, but increase with values of size parameter and refractive index. To stablish the full analogy with the scattering resonances, it is convenient to introduce the partial internal efficiencies, similar to Eq. (1), as:

$$F_\ell^{(e)} = \frac{2(2\ell+1)}{q^2}|c_\ell|^2, \qquad F_\ell^{(m)} = \frac{2(2\ell+1)}{q^2}|d_\ell|^2, \tag{7}$$

and, for the general characterization of the field inside the particle, introduce:

$$F_{sca} = \sum_{\ell=1}^{\infty}\left(F_\ell^{(e)} + F_\ell^{(m)}\right). \tag{8}$$

In Fig. 2 we plot the characteristic positions of different resonances supported by a sphere with refractive index $n = 4$ as a function of its size parameter, as well as the maximal values of the corresponding amplitudes $Q_\ell$ and $F_\ell$. For $\ell \geq 3$, the resonances are very sharp and we show them in the right panel of the figure with higher resolution of the size parameter. Although the position of the resonances in both functions, $Q_{sca}$ and $F_{sca}$, are quite close, there is a huge difference (by a few order of magnitude) in the amplitudes.

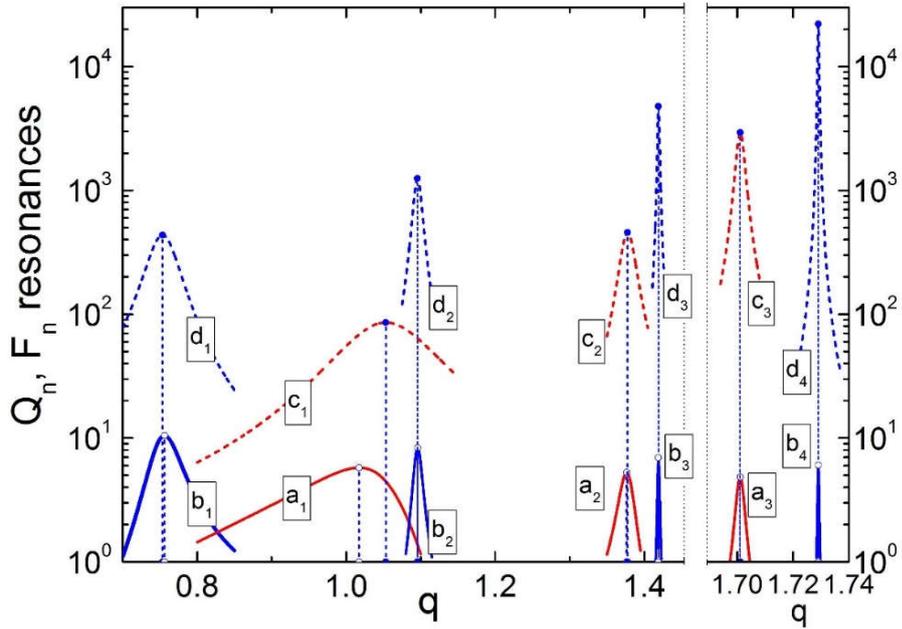

Fig. 2. Partial scattering efficiencies $Q_\ell$ and $F_\ell$, related to scattering amplitudes, $a_\ell$ and $b_\ell$, and internal field amplitudes, $c_\ell$ and $d_\ell$, for a spherical particle with refractive index $n = 4$ as a function of its size parameter $q$. Solid lines represent the partial scattering efficiencies $Q_\ell$, and dotted lines the internal partial efficiencies $F_\ell$. Electric amplitudes are shown in red color and magnetic amplitudes in blue. Note the change in the x-axis scale, introduced to present the sharp resonances found for $\ell \geq 3$.

Even more interesting is to analyse the maximal values of the internal fields at each of these resonances. These maximal values for the magnetic intensities are shown in Fig. 3 for each of the magnetic resonances excited in the particle. These high index, non-dissipative spherical particles, presents high quality factor (Q) resonances, for which the field values are very high (e.g. $H^2 \approx 8000$ for the octupole resonance). Note that the maximum value of the magnetic intensity increases and order of magnitude with each increase of resonance order $\ell$. In this way, $H^2_{max}(d_4) \approx 10\, H^2_{max}(d_3) \approx 10^2\, H^2_{max}(d_2) \approx 10^3\, H^2_{max}(d_1)$.

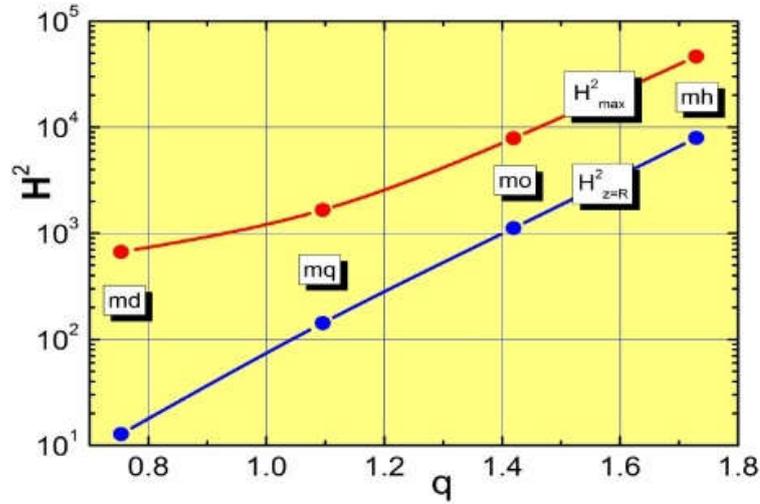

Fig. 3. Maximal values of $H^2$ (normalized to the incidence ones) inside the particle (red dots) for the first four internal magnetic resonances. The intensity of the magnetic field at the particle surface, in the point opposite to the wave incidence, are also shown (blue dots). Smooth solid lines are shown just for better visibility. Particle refractive index $n = 4$.

The electric and magnetic field distributions inside the particle (normalized to the incidence ones) and in the near field region along the z-axis (incidence one), passing through the particle centre are shown in Fig. 4. One can see that the fields inside the particle reach values significantly larger than those at the particle border. In Fig. 5, we plot the more detailed 2D distributions of the internal electric and magnetic fields (within the {x, z} and {y, z} cross-section planes) for the first three electric and magnetic resonances (dipolar, quadrupolar and octupolar). Note the large magnetic field enhancement obtained inside the particle, which is significantly larger than that of the electric field (e.g. around 30 times at the magnetic dipole resonance). Note also the formation of whispery-gallery-mode-like[41,42] field distributions for high-order multipolar resonances – specifically, they can be seen for as low order as the octupole (even quadrupole) in the {x, z} projection for the magnetic resonance and in the {y, z} projection for electric one.

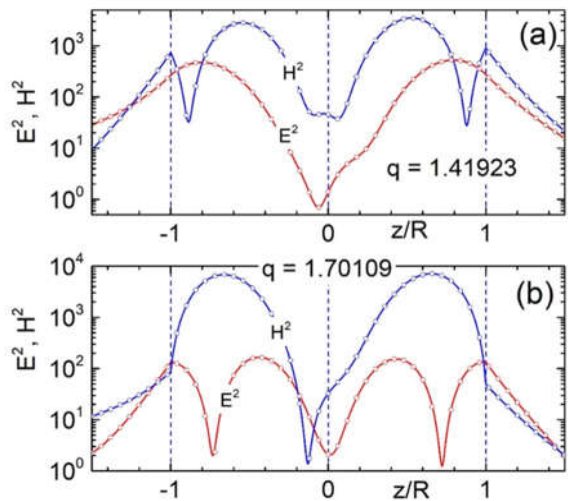

Fig. 4. Electric and magnetic field distributions for a particle with refractive index $n = 4$ and size parameter (a) $q = 1.41923$ (magnetic octupole resonance) and (b) $q = 1.70109$ (electric octupole resonance).

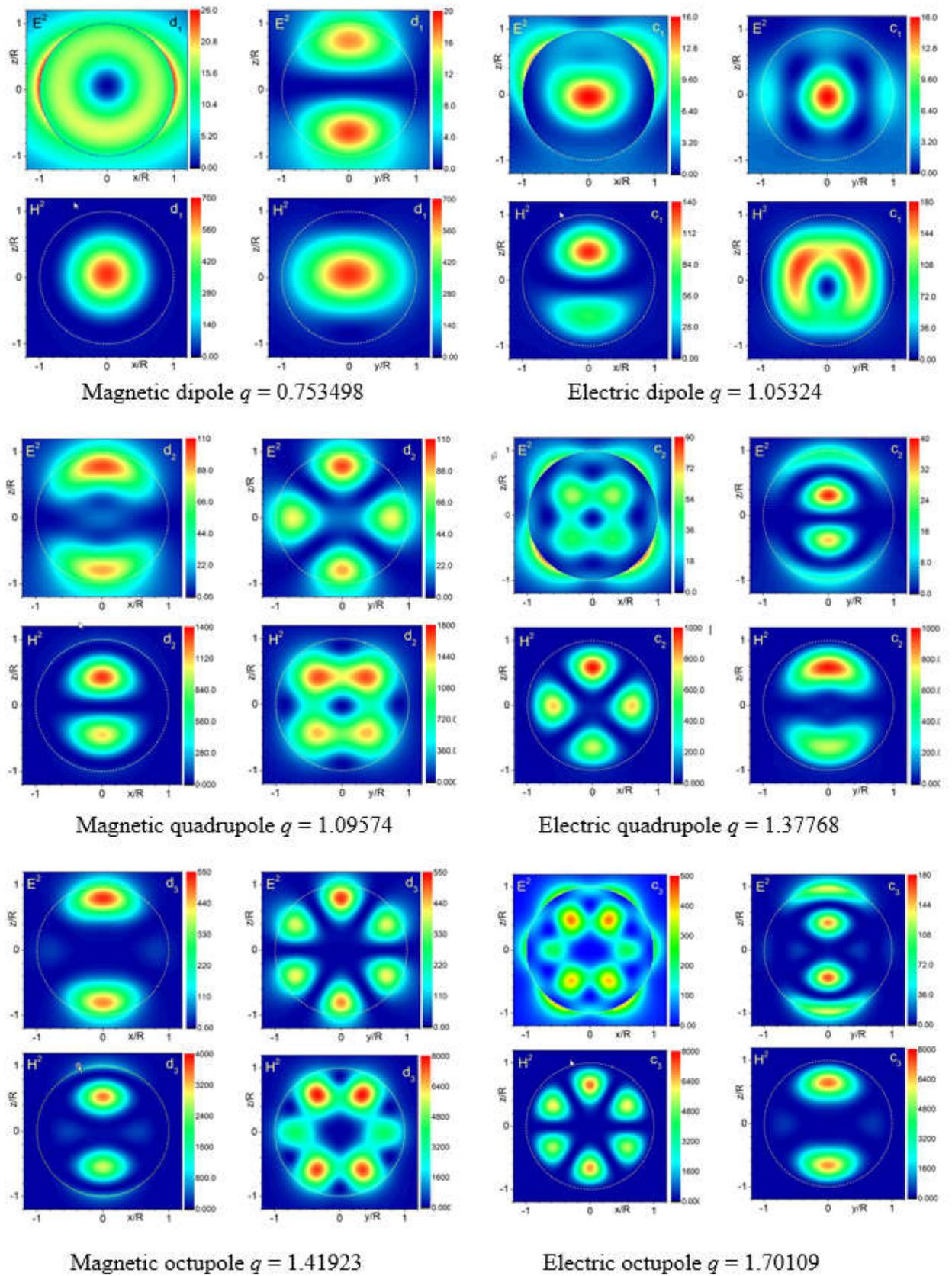

Fig. 5. Distribution of the internal electric and magnetic fields for dipole, quadrupole and octupole resonances.

When plotting the maximum field enhancement on an extended range of size parameters, an interesting phenomenon can be observed. For certain values of the size parameter, the excited resonance has a much larger amplitude than the rest of the resonances around it. As an example, for a particle with $n = 4$ this happens for $q \approx 19.415$, for which electric field enhancement as large as 20000 is obtained (see Fig.6a). This occurs near the first caustic in the paraxial approximation, $z/R = 1/(n-1) = 1/3$. Similar resonances can be seen at any value of refractive index, e.g. $n = 2$ or $n = 1.5$. Resonance can be also identified by plotting the fields at $z = R$. In Fig. 6b, we plot a large range of size parameters for the case of $n = 2$. An excited resonance having a much larger amplitude than the rest of the resonances around it can be clearly seen. That is the case of the resonance highlighted in the plot by an arrow, and to which we refer as super-resonance.

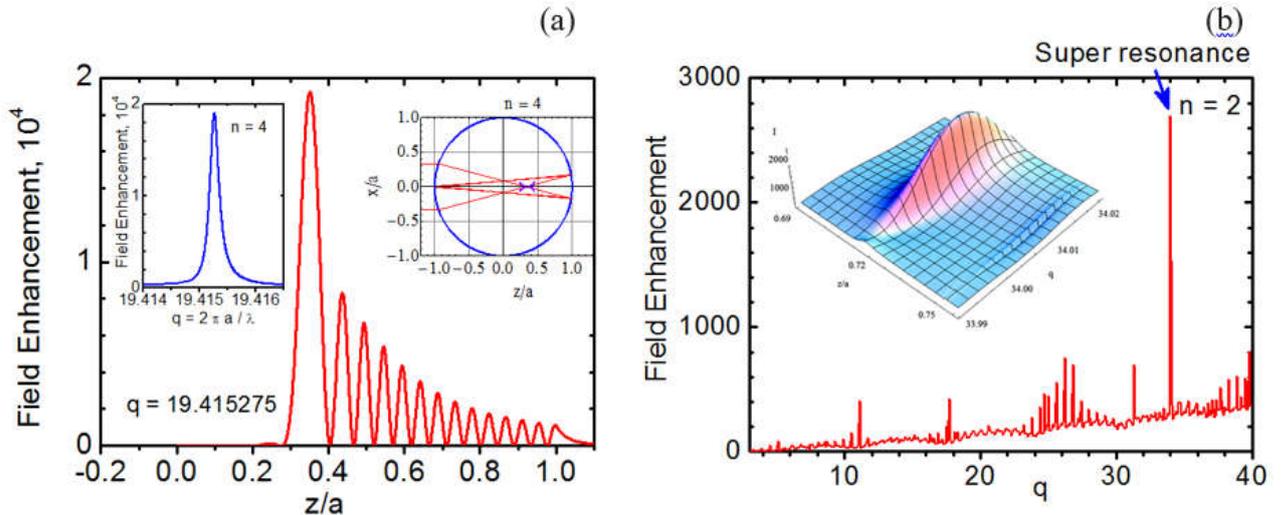

Fig. 6. (a) Distribution of field enhancement as a function of the position (normalized to the particle radius) inside the particle with $n = 4$. Field is almost symmetrical for $z < 0$, thus, we show just the part of the distribution with $z > -0.2$. The left inset shows the field vs size parameter at the point of $z/R \approx 1/3$. The right inset shows the ray tracing under the geometrical optics approximation. Caustics from transmitted light and caustics formed by the third reflection coincide at $n = 4$: the two edges meet each other! (b) Maximum field enhancement inside a particle with refractive index n = 2 as a function of the size parameter, showing the formation of a super-resonance (as indicated by an arrow). The left inset shows a zoomed-in view of the super-resonance.

Within the context of Mie theory, we can predict the spectral position and amplitude of the super-resonances for each multipolar order $\ell$, by finding the global maximum of the corresponding internal field efficiencies $F_\ell^{(e)}$ and $F_\ell^{(m)}$. In Fig. 7 we show the behaviour of these efficiencies as a function of the size parameter for a particle with a refractive index $n = 1.5$ and the multipolar order $\ell = 5$. The inset in Fig. 7 shows the positions of the global maxima of the resonances as a function of the multipolar order.

One of the particularities of super resonances are the peculiar field distributions associated with them. An example of such field distributions is plotted in Fig. 8a. In the figure, one can see two "bright electric points" within the particle, associated with very large field intensities (around 40 thousands) in the *x-z* plane. The presented field distributions correspond to a particle with refractive index $n = 1.5$ at the super resonance condition of the $\ell = 35$ multipolar order, excited for a size parameter $q = 26.9416$. In *y-z* plane (Fig. 8b), one can see the clear signature of a whispering gallery mode. Maximal magnetic field enhancements of the order of $10^4$-$10^5$ are also found at the corresponding super resonance conditions (Fig. 8c and Fig. 8d).

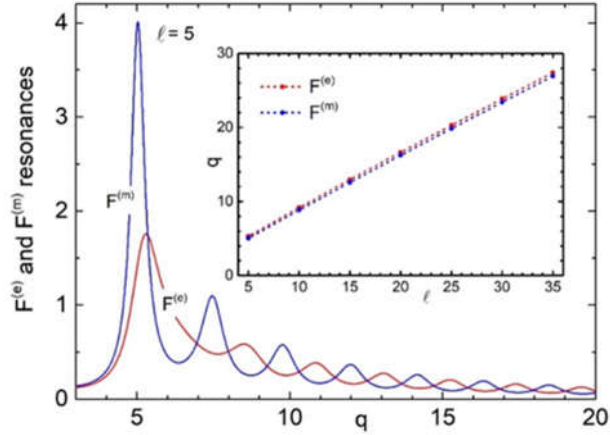

Fig. 7. Internal field efficiencies $F_\ell^{(e)}$ and $F_\ell^{(m)}$ as a function of the size parameter for the $\ell = 5$ multipolar orders for a particle with a refractive index $n = 1.5$. The inset shows the positions of the global maxima of the resonances as a function of the multipolar order.

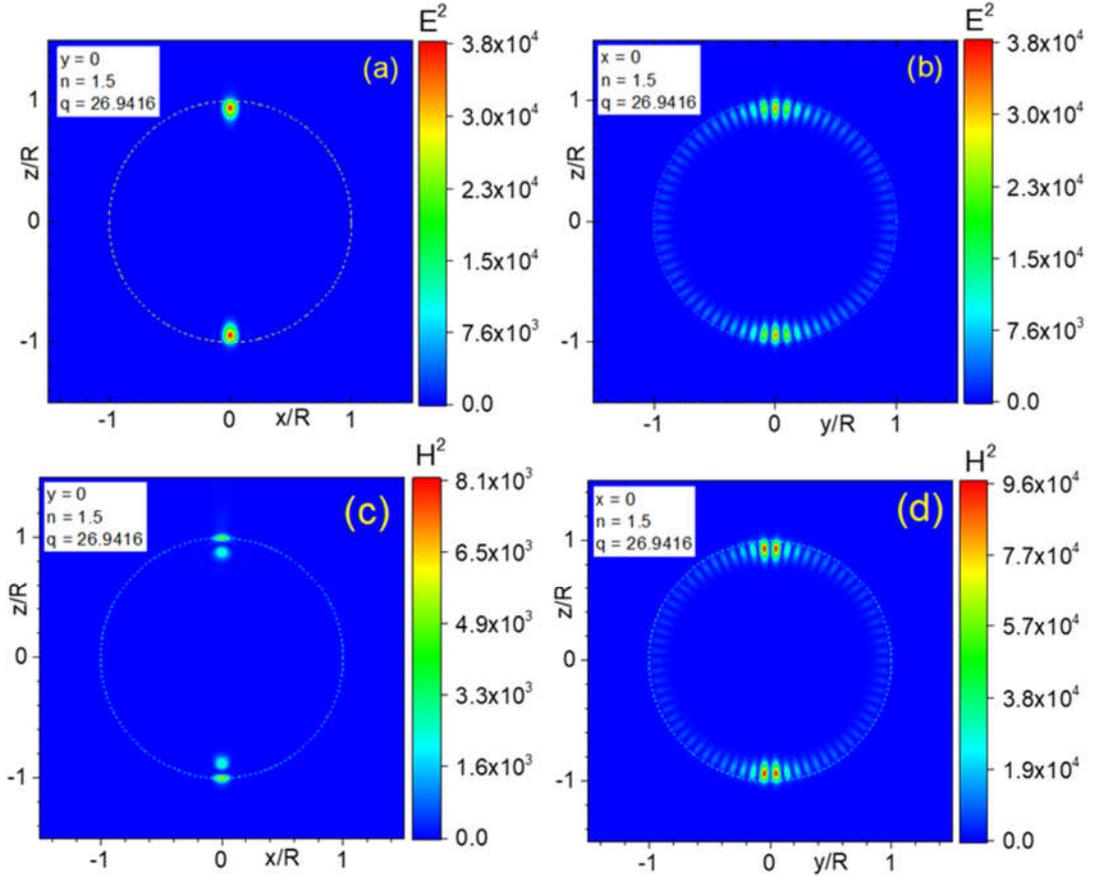

Fig. 8. Distributions of electric (a) (b) and magnetic (c) (d) fields in the *x-z* plane (a), (c) and in the *y-z* plane (b), (d) for a particle with refractive index $n = 1.5$ at the super resonance mode corresponding to the multipolar order $\ell = 35$.

It is important to note that this extremely large field enhancement is caused by a single resonant mode with high Q-factor. In Fig. 9a we illustrate the contribution of this resonance mode to the total distributions of electric and magnetic fields. It can be seen that the maximal field enhancement from this resonant mode is even higher than the total field, due to destructive interference with the other, non-resonant modes. In Fig. 9b we illustrate the high degree of localization of electric and magnetic fields inside the particle, beyond the diffraction limit.

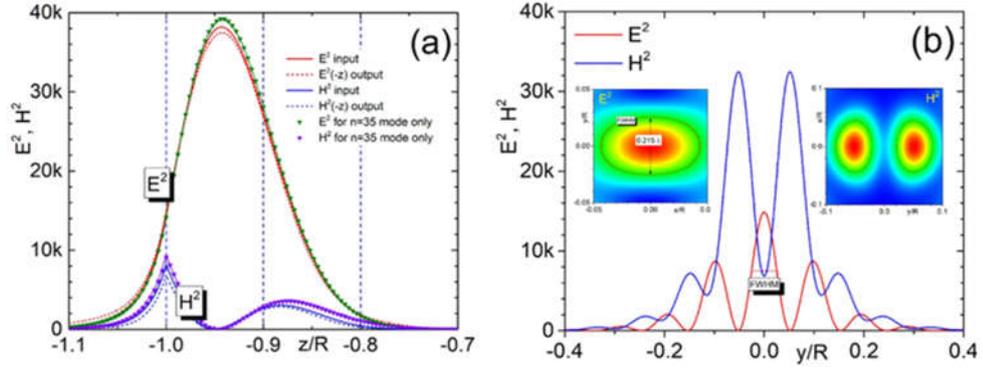

Fig. 9. (a) Distribution of the electric $E^2$ and magnetic $H^2$ fields along the $z$-axis passing through the particle centre and near the particle edges (where the field intensity localizes). The refractive index is $n = 1.5$ and the size parameter $q = 26.9416$, corresponding to the $\ell = 35$ multipolar super resonance. The electric field in $x$-$z$ plane is localized in two "bright points" near the input and the output windows on the surface of particle, as shown in Fig. 10a. The total electric and magnetic fields show some asymmetry while the field distributions for the single $\ell = 35$ multipolar contribution are symmetrical. (b) Distribution of the electric $E^2$ and magnetic $H^2$ fields along the $y$-axis at the output plane at $z = R$. 2D field distributions within these planes are shown in the insets. The solid lines in the 2D insets indicate the position of the full width at half maximum (FWHM). One can see high light localization along the $y$-direction: for electric field the FWHM is $0.215\,\lambda$, less than the diffraction limit given by $\lambda/2n \approx 0.333\,\lambda$.

In Fig. 10 we present the maximum values of the electric and magnetic field intensities associated with super resonances for a particle with refractive index $n = 1.5$. In Fig. 10a we focus on showing the positions of such super resonances in the vicinity of the size parameters $q = 5, 10, 15, 20$ and 25. Super resonances for higher size parameters ($20 < q < 30$) are shown in Fig. 10b (in the case of electric resonances) and Fig. 10c (in the case of magnetic ones). Note the narrow character of the super resonances and that, within the given range of size parameters, the maximal amplitude of the super resonances increase linearly. Note also that, in the visible range of frequencies, particles with size parameter $q = 30$ are on the micrometer size range, $2R \cong 10\lambda$. Thus, for a 6 $\mu m$ glass particle it would be theoretically possible to obtain near a five orders of magnitude field enhancement (see in Fig. 10b and Fig. 10c), a value that, in practice, might be bounded by fabrications imperfections and by dissipation effects. A final note regarding the observed spectrum is the strongly asymmetric character of the resonances observed, exhibiting Fano lineshapes[15,16,43].

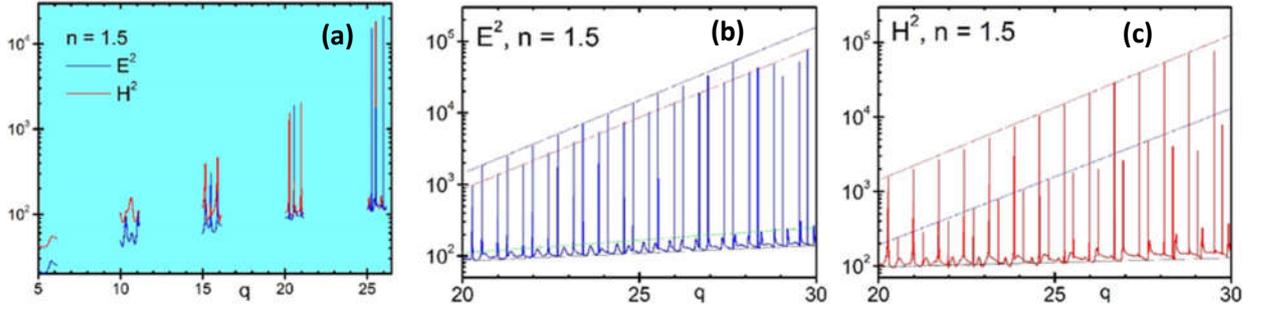

Fig. 10. The super resonances approximately size parameter $q = 5, 10, 15, 20$ and $25$. We show just these selected regions because it is not easy to distinguish electric (blue line) and magnetic (red lines) resonances due to their overlapping (a). Separately these resonances are shown in (b) and (c) for the region of high size parameter $20 < q < 30$. Resonances are very sharp.

As mentioned, dissipation may hinder some of the effects described here. For some materials with refractive index less than two (e.g. glass), this dissipation can be sufficiently small in the optical range as to still observe it. However, the role of dissipation effect for super resonances is not so clear for semiconductors with $n \geq 4$ (e.g. Si). Thus, we check the effect of dissipation for $n = 4$ and size parameter $q = 4.4241$, where the fields in non-dissipative material at super resonance reach value about $10^6$ (for $E^2$) and $10^7$ (for $H^2$), as shown in Fig. 11, whenever dissipate losses are included (even as low as $10^{-4}$) the field enhancement shows a drastic reduction. This demonstrates the high sensitivity of super resonances to dissipation.

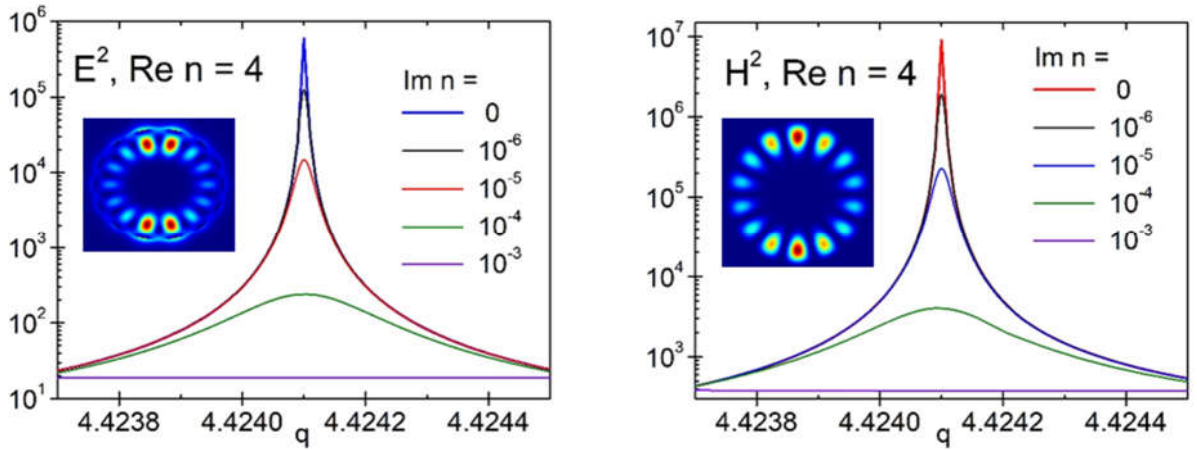

Fig. 11. Electric and magnetic field enhancement at super resonance for a particle with a complex refractive index $\tilde{n} = n + i\kappa$, where $n = 4$ and a few values of $\kappa \leq 10^{-3}$. Super resonances can be clearly seen for $\kappa \leq 10^{-5}$, while for dissipation values above, the enhancement drastically reduces. The insets show the corresponding 2D field distributions for the dissipation-less case.

In any case, theoretical estimations show the ability to reach enhancement of magnetic field at least on the level of $10^5$. Naturally, the magnetic field of light is small (e.g. a laser with an intensity $I = 5 \cdot 10^{16}$ W/cm² yields a magnetic induction in vacuum of about 3 T). Within a dielectric particle supporting a super resonance, a field enhancement of four orders of magnitude could potentially allow to reach magnetic induction above 300 T (exceeding the magnetic field of a typical White dwarf star). Using a $5 \cdot 10^{18}$ W/cm² fs laser one could exceed the Human record for pulsed magnetic field, 2800 T[44]. It is sufficiently close to interatomic magnetic fields, on the order of $10^5$

T. It means that with super resonances in dielectrics, it might be possible to realize magnetization-induced optical nonlinearity. Until now, nonlinear effects in magnetization-induced optical nonlinearity were observed mainly in thin ferromagnetic films[45,46].

Beyond the ability to create highly localized fields, both inside the particle and outside in the near field region, there are a few reasons to investigate such super resonances. For example, they might open new venues in many modern applications such as in photonic nanojet generation[14], white-light superlens nanoscopy[22] and surface-enhanced Raman spectroscopy (SERS)[47]. Also, the might serve as a means for optimization of whispering gallery modes, widely used for telecommunication applications in, e.g., wavelength division multiplexing (WDM) schemes[41,42].

In conclusion, we reveal novel, super-resonance modes supported by dielectric spheres. These resonances, present for all multipolar orders, exhibit Fano lineshapes and have extraordinarily high associated electric and magnetic field enhancements, which increase linearly with the multipolar order. The phenomenon can be observed at visible frequencies using simple glass microspheres, which may allow enhancing the magnetic field of light (which is typically small) to a few order of magnitudes. We believe these super-resonances are an attractive platform for some promising applications, like e.g. enhanced absorption effect, ablation caused by magnetic pressure, or those mentioned above.